\documentclass[prb,showpacs,preprintnumbers,amsfonts,amssymb,amsmath,floats,twocolumn,superscriptaddress,aps]{revtex4}

\usepackage{graphicx}
\usepackage{dcolumn}
\usepackage{bm}
\usepackage[final,dvips]{epsfig}
\usepackage{bm}
\usepackage{color}

\newcommand{\be}{\begin{equation}}
\newcommand{\ee}{\end{equation}}
\newcommand{\ba}{\begin{eqnarray}}
\newcommand{\ea}{\end{eqnarray}}
\newcommand{\ff}[1]{{\bm #1}}
\newcommand{\tr}{\mbox{tr}}
\newcommand{\Tr}{\mbox{Tr}}

\begin{document} 

\title{ 
Antiferromagnetic to superconducting phase transition in the 
hole- and electron-doped Hubbard model at zero temperature
} 

\author{M. Aichhorn}

\affiliation{
Institute for Theoretical Physics, University of
W\"urzburg, Am Hubland, 97074~W\"urzburg, Germany
}

\author{E. Arrigoni}

\affiliation{
Institute for Theoretical Physics and Computational
Physics, Graz University of Technology, Petersgasse 16, 8010
Graz, Austria
}

\author{M. Potthoff}

\affiliation{
Institute for Theoretical Physics, University of
W\"urzburg, Am Hubland, 97074~W\"urzburg, Germany
}
 
\author{W. Hanke}

\affiliation{
Institute for Theoretical Physics, University of
W\"urzburg, Am Hubland, 97074~W\"urzburg, Germany
}
 
\begin{abstract}
The competition between $d$-wave superconductivity (SC) and 
antiferromagnetism (AF) in the high-$T_c$ cuprates is investigated
by studying the hole- and electron-doped two-dimensional Hubbard model 
with a recently proposed variational quantum-cluster theory. 
The approach is shown to provide a thermodynamically consistent 
determination of the particle number, provided that an overall shift of
the on-site energies is treated as a variational parameter.
The consequences for the single-particle excitation spectra and for the  
phase diagram are explored.
By comparing the single-particle spectra with quantum Monte-Carlo (QMC) 
and experimental data, we verify that the low-energy excitations in a 
strongly-correlated electronic system are described appropriately. 
The cluster calculations also reproduce the overall ground-state phase diagram 
of the high-temperature superconductors. 
In particular, they include salient features such as the enhanced robustness 
of the antiferromagnetic state as a function of electron doping and the 
tendency towards phase separation into a mixed 
antiferromagnetic-superconducting phase at low-doping and a pure 
superconducting phase at high (both hole and electron) doping.
\end{abstract} 
 

\maketitle

\section{Introduction}

The theory of high-temperature superconductivity remains one of the
most challenging problems in solid-state physics. In many metallic
compounds, over a broad range of compositions and temperatures, only 
two phases are encountered, i.e.\ the normal (Fermi liquid) and a
magnetic or superconducting phase.
In sharp contrast, the high-temperature superconductors (HTSC) and other 
strongly correlated electron systems, such as heavy-fermion and a variety 
of transition-metal oxide systems, exhibit many ordered phases, which 
appear to compete and sometimes coexist.\cite{im.fu.98} 
In the HTSC, besides ordered antiferromagnetic (AF) and superconducting (SC) 
phases, compelling evidence exists for charge- and spin-``stripe'' phases and 
phases with coexisting SC and AF order.\cite{kive} 
The common denominator and the underlying reason for these competing orders 
is certainly the presence of strong electronic correlations and Mott-Hubbard 
physics: 
The interplay between kinetic-energy and Coulomb-correlation effects induces 
an extreme sensitivity to external parameters (doping, temperature, pressure,
etc.) and a rather difficult to predict ``outcome'', i.e.\ the characteristic  
low-energy excitations and the phase diagram at low temperatures.
The central challenge in the field of high-$T_c$ superconductivity is, therefore, 
the connection of the (known) microscopic interactions at the level of
electrons and ions, which are at high energy ($\sim$eV) and temperature
$T$, with the ``emerging phenomena'' at $T=0$, i.e.\ competing and
nearly degenerate orders.
Ideally, one should employ a systematic renormalization-group 
approach to integrate out the irrelevant degrees of freedom and, thereby,
correctly bridge high to low energies and eventually go to $T=0$. 
It is, however, by no means obvious how to do this when
strong correlations are present, such as in the HTSC.

In this context cluster techniques, which systematically approach the
infinite-size (low-energy) limit, provide a powerful alternative. 
\cite{ma.ja.05,tr.ky.05u,ha.ai.05}
In this paper, we will discuss and apply a variational-cluster approach (VCA)
which was proposed recently. \cite{po.ai.03,da.ai} 
It is based on the self-energy-functional theory (SFT), \cite{pott.epj} which 
provides a general variational scheme to use dynamical information from an exactly 
solvable ``reference system'' (in our case, an isolated cluster) to go to the 
infinite-size lattice fermion problem at low temperatures and at $T=0$, in particular.
In our earlier work \cite{da.ai} this scheme was formulated to study phases with 
spontaneously broken symmetry. For the cluster sizes used, it was shown that the VCA 
correctly reproduces long-range AF order for the two-dimensional ($2D$) Hubbard
model and the absence of this order in one dimension. 
This non-trivial ``test'' implies that the VCA goes well beyond ordinary 
mean-field theory.

Another crucial test is provided by the dynamical information contained
in the one-particle Green's function $\ff G$. 
Compared to variational schemes based on wave functions, \cite{para} an important 
advantage of the VCA consists in the fact that it quite naturally gives the 
one-electron Green's function $\ff G$.
For the $2D$ Hubbard model, it was recently demonstrated \cite{da.ai} that the
VCA, with the lattice tiled by ($\sqrt{10} \times \sqrt{10}$) clusters, correctly
reproduces low-temperature quantum Monte-Carlo (QMC) data, in particular, the 
coherent and incoherent ``bands'' experimentally known from ARPES data. \cite{da.hu}

These tests provide also the foundation for attacking the question whether
the ``minimal'' microscopic model, namely the $2D$ one-band Hubbard
model, reproduces the essential features of the electron- and hole-doped
HTSC phase diagram. We will not go into a lengthy discussion of what
interactions should be retained at the electron-ion level. But, when
choosing the $2D$ one-band Hubbard model, \cite{ande} i.e.\
\begin{equation}
        H = \sum_{ij\sigma} t_{ij} c^{\dagger}_{i\sigma} c_{j\sigma} +
        U \sum_{i} n_{i \uparrow} n_{i \downarrow} \: ,
\label{eq:hub}
\end{equation}
where  $c_{j\sigma}$, $c^{\dagger}_{i\sigma}$ are the usual annihilation 
and creation operators, $t_{ij}$ denote the hopping-matrix elements,
$n_{i\sigma} = c_{i\sigma}^\dagger c_{i\sigma}$ the density at site $i$ 
with spin $\sigma=\uparrow,\downarrow$ 
and $U$ the local Coulomb repulsion, one has introduced gross simplifications, 
leaving out other orbital (e.g.\ $p$) degrees of freedom, long-range 
Coulomb interaction, electron-phonon coupling, etc. Nevertheless, 
this choice appears to be legitimate, last not least in view 
of the amazing agreement achieved between numerical simulations 
and experimental results for the normal-state properties of the 
cuprates (see, for example, Refs.\ \onlinecite{im.fu.98,ande,pr.ha.95,pr.ha.97}).

The ground-state phase diagram of the model was recently investigated
using the VCA by S\'{e}n\'{e}chal et al.\ \cite{se.la.05}
and by two of us. \cite{ai.ar.05} 
There are important technical differences, but the ``upshot'' of the two works 
is as follows: 
For the cluster sizes used in the VCA, the $T=0$ phase diagram of the Hubbard model 
(including hopping terms up to second or third-nearest neighbors) turns out
to be qualitatively similar to that of the electron- and hole-doped cuprates. 
The model correctly describes the overall phase diagram, such as the occurrence 
of the AF and SC phases and predicts the corresponding doping ranges in qualitative
agreement with the experiments for the cuprate materials.

The present paper has several purposes:
First, we would like to stress that for an application of the VCA to the 
high-$T_c$ problem it is of crucial importance to treat the on-site
energies in the reference system as variational parameters. 
We will show that this ensures a thermodynamically consistent 
determination of the average particle number.
Compared to the study of Ref.~\onlinecite{se.la.05} this 
represents an important methodical extension.
On the other hand, 
without the variational optimization of the on-site energy, one
has to tolerate an inconsistency in the determination of the average particle number. 
The effects of this error shall be demonstrated by model calculations.
The issue of thermodynamic consistency is also discussed for the
off-diagonal elements of the one-particle density matrix and, in case of 
spontaneous symmetry breaking, for the respective order parameter.
It is interesting to note that there are no such problems in the dynamical 
mean-field theory (DMFT) \cite{me.vo.89,ge.ko.96} and its cluster extensions. 
\cite{ko.sa,he.ta.98}
Here the on-site energies are kept at their ``physical'' values from 
the very beginning rather than being determined from the self-consistency
condition.
The (cellular) DMFT, however, can be considered as a special approximation 
within the general SFT framework. \cite{po.ai.03,da.ai}
Hence, the question arises why a fixed on-site energy does {\em not} spoil 
thermodynamic consistency in the case of (cellular) DMFT.
It is interesting to note that, in case of interacting bosons, the issue of 
consistency has also been shown to be very important, recently. \cite{ko.du.05}

Second, an accurate analysis of the behavior of the chemical potential
as a function of the particle density close to the transition to
a non-magnetic state, as well as a corresponding Maxwell construction,
indicate the presence of an inhomogeneous ground state with macroscopically 
large regions of low- and high-particle density.
Using $L_c=4$-site clusters, we can get a rough estimate of the ground-state 
phase diagram and investigate the instability of the homogeneous (AF, SC) 
phases against charge inhomogeneities.
This represents an important complement to the work of Ref.\ \onlinecite{se.la.05},
where larger cluster sizes up to $L_c=10$ sites have been considered but without 
an appropriate analysis via a Maxwell construction.

Finally, our numerical results give valuable insights into different questions 
of the high-$T_c$ problem, as they provide direct access to the single-particle 
excitation spectrum in the strong-coupling regime at zero temperature.
This has to be contrasted with the work of Maier et al.\ \cite{ma.ja.05u,ma.ja.05u.ps} 
who have been able to treat clusters with $L_c > 20$ sites within the dynamical 
cluster approximation (DCA), but are restricted to finite temperatures, intermediate 
coupling $U/t \lesssim 4$ and imaginary time. 
A review comparing the application of different cluster
(SFT, DCA, and cellular DMFT) as well as weak-coupling methods for the
Hubbard model has appeared recently. \cite{tr.ky.05u}

Our paper is organized as follows: 
We start with a summary of the central ideas of the SFT in general (Sec.\ \ref{sec:sft}).
Thermodynamic consistency with respect to the average particle number is discussed
in Sec.\ \ref{sec:n}.
Consistency with respect to the off-diagonal elements of the density matrix is addressed 
in Appendix \ref{sec:cicj} and the case of the C-DMFT is discussed in Appendix \ref{sec:cdmft}.
Our theoretical considerations are completed by describing some computational 
details of the VCA calculations in Sec.\ \ref{sec:vca}.
Section \ref{spe} then presents the results for the $T=0$ spectral function
$A(\ff{k},\omega)$ for electron- and hole-doping. 
We discuss how the characteristically different doping dependencies of the 
spectra give rise to different Fermi-surface evolutions upon doping.
These Fermi-surface evolutions can then be tied up with the
characteristic differences in the electron- and hole-doped phase diagrams, 
such as the enhanced robustness of the AF order in the electron-doped case. 
The ground-state phase diagram will be presented and discussed in detail 
in Sec.\ \ref{gspd}.
Finally, Sec.\ \ref{summ} contains our main conclusions and a summary. 

\section{Theoretical background}
\label{sec:tb}

\subsection {Self-energy-functional theory}
\label{sec:sft}

The central idea of the self-energy-functional theory (SFT) \cite{pott.epj} is to 
make use of the universality of the Luttinger-Ward functional $\Phi_{\ff U}[\ff G]$ 
\cite{lu.wa.60} or of its Legendre transform $F_{\ff U} [\ff \Sigma]$:
For a system with Hamiltonian $H=H_{0} (\ff t) + H_{1} (\ff U)$, where $\ff t$ are 
the one-particle and $\ff U$ the interaction parameters, the functional dependence
$\Phi_{\ff U}[\cdots]$ or $F_{\ff U}[\cdots]$ is independent of $\ff t$.
This universality is obvious as the Luttinger-Ward functional is defined via 
a skeleton-diagram expansion involving dressed propagators and vertices only.
\cite{lu.wa.60}

Concentrating on the self-energy $\ff \Sigma$ instead of the single-particle
Green's function $\ff G$, the grand potential of the system at temperature $T$
and chemical potential $\mu$ can be written as a functional of $\ff \Sigma$: 
\begin{equation}
        \Omega_{\ff t, \ff U} [\ff \Sigma] =  \Tr \ln (\ff G_{0, \ff t}^{-1} -
                \ff \Sigma)^{-1} + F_{\ff U} [\ff \Sigma] \: ,
\label{eq:sf}           
\end{equation}
where $\ff G_{0,\ff t} = (\omega + \mu - \ff t)^{-1}$ is the free Green's function
and $\mbox{Tr} \equiv T \sum_{\omega_n} e^{i\omega_n 0^+} \tr$ with the usual trace
$\tr$ and the Matsubara frequencies $\omega_n = (2n+1)\pi T$ for integer $n$.
At the physical self-energy $\ff \Sigma = \ff \Sigma_{\ff t,\ff U}$, 
the grand potential is stationary: $\delta \Omega_{\ff t, \ff U} 
[\ff \Sigma_{\ff t , \ff U}] = 0$.

Why is it more advantageous to express $\Omega$ as a functional of the
self-energy $\ff \Sigma$ rather than $\ff G$? This has to do with the
``short-range'' character of $\ff \Sigma$ as a function of its real-space 
coordinates, \cite{ha.sh} which in general is due to the fact that 
$\ff \Sigma$ is qualitatively related to a dynamically screened particle-particle 
interaction. 
For the Hubbard model, in particular, one believes that important effects
are sufficiently accounted for by a local \cite{mull} or short-ranged
\cite{sc.cz} self-energy, at least for high lattice dimensions.
A local or short-ranged self-energy, however, can be well generated by
a cluster of finite size, and for the subsequent optimization of 
the cluster trial self-energy, the self-energy functional (\ref{eq:sf})
can be used.
This concept allows for the construction of a class of conceptually 
clear and thermodynamically consistent approximations,
including the dynamical mean-field theory (DMFT) \cite{me.vo.89,ge.ko.96} and a 
cluster extension of the DMFT \cite{ko.sa} (see Ref.\ \onlinecite{pott.ass} 
for a detailed discussion).

Due to the universality of $F_{\ff U}[\ff \Sigma]$, we have 
\begin{equation}
        \Omega_{\ff t', \ff U} [\ff \Sigma] = F_{\ff U} [\ff \Sigma] + \Tr \ln
        (\ff G_{0,\ff t'}^{-1}-\ff \Sigma)^{-1}
\label{eq:sfp}
\end{equation}
for the self-energy functional of a so-called ``reference system'', which
is given by a Hamiltonian with the same interaction part $\ff U$ but modified
one-particle parameters $\ff t'$: $H'=H_{0} (\ff t') + H_{1} (\ff U)$.
Although it has different microscopic parameters, 
the reference system is assumed to be in the same macroscopic state as the 
original system, so it has the same temperature $T$ and the same chemical 
potential $\mu$.
By a proper choice of its one-particle part, the problem posed by the reference 
system $H'$ can be much simpler than the original problem posed by $H$, 
such that the self-energy of the 
reference system, $\ff \Sigma_{\ff t',\ff U}$, can be computed exactly within a
certain subspace of parameters $\ff t'$.
Combining Eqs.\ (\ref{eq:sf}) and (\ref{eq:sfp}), we can eliminate the functional
$F_{\ff U}[\ff \Sigma]$.
Inserting as a trial self-energy the self-energy of the reference
system, then
yields:
\begin{equation}
        \Omega_{\ff t, \ff U} [\ff \Sigma_{\ff t',\ff U}] 
        = 
        \Omega_{\ff t', \ff U}
        +
        \Tr \ln (\ff G_{0,\ff t}^{-1}-\ff \Sigma_{\ff t',\ff U})^{-1}
        - 
        \Tr \ln \ff G_{\ff t',\ff U} \: ,
\label{eq:ocalc}        
\end{equation}
where $\Omega_{\ff t', \ff U}$ and $\ff G_{\ff t',\ff U}$ are the grand
potential and the Green's function of the reference system.
Stationary points are obtained, and this is the approximation,
by restricting the variation to the subspace of trial self-energies 
$\ff \Sigma_{\ff t',\ff U}$:
\begin{equation}
  \frac{\partial \Omega_{\ff t, \ff U} [\ff \Sigma_{\ff t',\ff U}]}
  {\partial \ff t'} = 0 \qquad \mbox{for} \quad \ff t' = \ff t'_{\rm s} \: .
\label{eq:stat}  
\end{equation}  
Varying the trial self-energy
means to vary the one-particle parameters $\ff t'$ of the reference system.
For further details of the approach see Ref.\ \onlinecite{pott.epj}.

Here, we will focus on the Hubbard model, Eq.\ (\ref{eq:hub}), as the original model
given by $H$.
Different possible choices for $H'$ and the corresponding systematics of 
approximations generated in this way are discussed in Ref.\ \onlinecite{pott.ass}.
In the following, we will concentrate on two cluster approaches: (i) the variational
cluster approach (VCA) \cite{po.ai.03,da.ai} and (ii) the cellular DMFT (C-DMFT).
\cite{ko.sa} 
The VCA can be seen as a variational generalization of the cluster-perturbation 
theory. \cite{gr.va.93,se.pe.00}
It is obtained by partitioning the infinite lattice into disconnected (identical) 
clusters of $L_c$ sites each and choosing $H'$ to consist of the intra-cluster parts 
only, i.e.\ the inter-cluster hopping is switched off in $H'$.
The C-DMFT is a cluster variant of the DMFT.
In the context of the SFT, it is obtained in the same way as the VCA but with
an additional coupling of each of the $L_c$ correlated cluster sites to a continuous 
bath, i.e.\ to an infinite number of uncorrelated additional bath
sites. The on-site energies of the bath sites as well as their coupling to the original 
sites are treated as variational parameters.

\subsection{Consistent determination of the particle density}
\label{sec:n}

Once a reference system is specified, one should, in principle, vary {\em all}
one-particle parameters of $H'$. 
This procedure would give the optimal result but requires a search for a stationary
point in a high-dimensional parameter space.
From a pragmatic point of view it is thus advisable to concentrate on a few
parameters only which have to be selected by physical arguments. 
Here, we argue that the variation of the on-site energies is important to achieve 
thermodynamic consistency with respect to the average particle number. 
In case of the Hubbard model, this means to consider the site-independent energy 
$\varepsilon' \equiv t'_{ii}$ as (one of the) variational parameter(s).

The average particle number $\langle N \rangle$ can be calculated in two
different ways:
on the ``zero-particle level'' by differentiation of the grand potential 
with respect to the chemical potential $\mu$:
\begin{equation}
  \langle N \rangle = - \frac{\partial \Omega}{\partial \mu} \: ,
\label{eq:n1}
\end{equation}  
and on the ``one-particle level'' by frequency integration of the one-particle
excitation spectrum:
\begin{equation}
  \langle N \rangle = \sum_{i\sigma} \langle n_{i\sigma} \rangle
  = \sum_{i\sigma} \int_{-\infty}^\infty f(\omega) A_{ii\sigma}(\omega) d\omega \: .
\label{eq:n2}
\end{equation}  
Here, $A_{ii\sigma}(\omega) = - \mbox{Im}\, G_{ii\sigma}(\omega + i 0^+) / \pi$ is
the local (possibly spin-dependent) spectral density with $\ff G = \ff G_{\ff t, \ff U}$
for short, and 
$f(\omega) = (\exp(\omega/T)+1)^{-1}$ is the Fermi function.
For simplicity, we exclude non-collinear magnetic states and assume all expectation
values to be diagonal in the spin index.

Thermodynamic consistency means that both ways of calculating $\langle N \rangle$
yield the same result.
Since $\Omega \equiv \Omega_{\ff t, \ff U} [\ff \Sigma_{\ff t'_{\rm s},\ff U}]$ is the 
{\em approximate} SFT grand potential at the stationary point $\ff t' = \ff t'_{\rm s}$,
and since the spectral density or, equivalently, the one-particle Green's function is 
the {\em approximate} Green's function given by 
$\ff G \equiv 1 / (\ff G_{0,\ff t}^{-1} - \ff \Sigma_{\ff t'_{\rm s},\ff U})^{-1}$,
the equivalence of (\ref{eq:n1}) and (\ref{eq:n2}) is by no means understood {\em a priori}.

To prove thermodynamic consistency, we start from Eq.\ (\ref{eq:n1}).
According to Eq.\ (\ref{eq:ocalc}), there is a twofold $\mu$ dependence of 
$\Omega=\Omega_{\ff t, \ff U} [\ff \Sigma_{\ff t'_{\rm s},\ff U}]$:
(i) an {\em explicit} $\mu$ dependence due to the chemical potential in the free 
Green's function of the original model, $\ff G_{0,\ff t}^{-1} = \omega + \mu - \ff t$,
and (ii) an {\em implicit} $\mu$ dependence due to the $\mu$ dependence of the self-energy
$\ff \Sigma_{\ff t'_{\rm s},\ff U}$, the Green's function $\ff G_{\ff t'_{\rm s},\ff U}$
and the grand potential $\Omega_{\ff t'_{\rm s},\ff U}$ of the reference system:
\begin{equation}
  \langle N \rangle 
  = 
  - \frac{\partial \Omega}{\partial \mu_{\rm ex.}} 
  - \frac{\partial \Omega}{\partial \mu_{\rm im.}} \: .
\end{equation}
Note that the implicit $\mu$ dependence is due to the chemical potential of the
reference system which, by construction, is in the same macroscopic state as the original 
system (with the same temperature $T$ and the same chemical potential $\mu$) 
{\em as well as} due to the $\mu$ dependence of the stationary point $\ff t'_{\rm s}$ 
itself.
This is a subtlety which, however, can be ignored since
\begin{equation}
  \frac{\partial \Omega}{\partial \ff t'} 
  \cdot
  \frac{\partial \ff t'}{\partial \mu} = 0
\end{equation}
for $\ff t' = \ff t'_{\rm s}$ because of stationarity condition (\ref{eq:stat}).
(Actually, only those elements of $\ff t'_{\rm s}$ show up a $\mu$ dependence
that are treated as variational parameters. 
According to the chain rule, however, the derivative of $\Omega$ has to be 
performed just with respect to those elements, with a vanishing result due to
the stationarity condition.)

The self-energy, the Green's function and the grand potential of the
reference system are defined as grand-canonical averages.
Hence, their $\mu$ dependence due to the grand-canonical Hamiltonian 
${\cal H'} = H' - \mu N$ is (apart from the sign) the same as their 
dependence on $\varepsilon'$:
$\partial/\partial \mu_{\rm im.} = - \partial/\partial \varepsilon'$, etc.
Consequently, we have:
\begin{equation}
  \langle N \rangle 
  = 
  - \frac{\partial \Omega}{\partial \mu_{\rm ex.}} 
  + \frac{\partial \Omega}{\partial \varepsilon'}
\: .
\label{eq:exim}
\end{equation}
The first derivative is readily calculated:
\begin{eqnarray}
  \frac{\partial \Omega}{\partial \mu_{\rm ex.}} \! &=&
  \frac{\partial}{\partial \mu_{\rm ex.}}
  T \sum_{\omega_n} e^{i\omega_n 0^+} 
  {\rm tr} \ln \frac{1}{ \ff G_{0,\ff t}^{-1}(i\omega_n) - \ff \Sigma_{\ff t'_{\rm s},\ff U}(i\omega_n)}
\nonumber \\ &=&
  T \sum_{\omega_n} e^{i\omega_n 0^+}
  {\rm tr} \, \frac{-1}{i\omega_n + \mu - \ff t - \ff \Sigma_{\ff t'_{\rm s},\ff U}(i\omega_n)}
\nonumber \\ &=&
  \frac{-1}{2\pi i} \oint_{\cal C} e^{\omega 0^+} f(\omega) \: \tr
  \frac{-1}{\omega + \mu - \ff t - \ff \Sigma_{\ff t'_{\rm s},\ff U}(\omega)}
  d\omega 
  \: .
\nonumber \\ 
\end{eqnarray}
Here, the contour $\cal C$ encloses the first-order poles of the Fermi function 
at $\omega_n=(2n+1)\pi T$ in counter-clockwise direction. 
Using Cauchy's theorem, we can proceed to an integration over real
frequencies.
Inserting into Eq.\ (\ref{eq:exim}), we get:
\begin{equation}
  \langle N \rangle 
   = 
  - \frac{1}{\pi} \mbox{Im}
  \int_{-\infty}^\infty f(\omega) \: \tr
  \frac{1}{\ff G_{0,\ff t}^{-1}
  - \ff \Sigma_{\ff t',\ff U}} \Bigg|_{\omega + i 0^+} d\omega 
  + \frac{\partial \Omega}{\partial \varepsilon'}
  \: 
\label{eq:nn}  
\end{equation}
for $\ff t' = \ff t'_{\rm s}$.

The first term on the right-hand side is just the expression for 
the average particle number given by (\ref{eq:n2}).
The second term on the right-hand side vanishes {\em provided that} the
variational condition (\ref{eq:stat}) is satisfied, i.e.\ {\em provided that}
$\varepsilon'$ is included in the set of variational parameters.
In this case one has thermodynamic consistency.
If $\varepsilon'$ was not treated as a variational parameter but kept at the
value given by the original system, $\varepsilon' = \varepsilon$, one would have
a finite $\partial \Omega / \partial \varepsilon'$ in (\ref{eq:nn}), and the two
expressions (\ref{eq:n1}) and (\ref{eq:n2}) for the average particle number would 
yield different results.
This completes the proof. 

Eq.\ (\ref{eq:n1}) for the average particle number of the lattice model
can be compared with 
\begin{equation}
   \langle N \rangle' = - \frac{\partial \Omega'}{\partial \mu} =
   \sum_{i\sigma} \int_{-\infty}^\infty f(\omega) A'_{ii\sigma}(\omega) \: d\omega 
   \: ,
\label{eq:nnp}  
\end{equation}
which gives the average particle number of the reference system. 
Again, there are two ways to get $\langle N \rangle'$: either as the derivative 
of the reference system's grand potential $\Omega'\equiv \Omega_{\ff t',\ff U}$ 
or by frequency integration of the reference system's spectral density 
$A'_{ii\sigma}(\omega) \equiv (-1/\pi) \mbox{Im} G'_{ii\sigma}(\omega+i0^+)$.
As the reference system is solved exactly, both ways must yield the same result.
Note, however, that $\langle N \rangle \ne \langle N \rangle'$ in general.

The above reasoning can straightforwardly be generalized to the off-diagonal
($i\ne j$) elements of the one-particle density matrix 
$\langle c_{i\sigma}^\dagger c_{j\sigma} \rangle$.
This is discussed in Appendix \ref{sec:cicj}.

\begin{figure}[t]
\begin{center}
  \includegraphics[width=0.7\columnwidth]{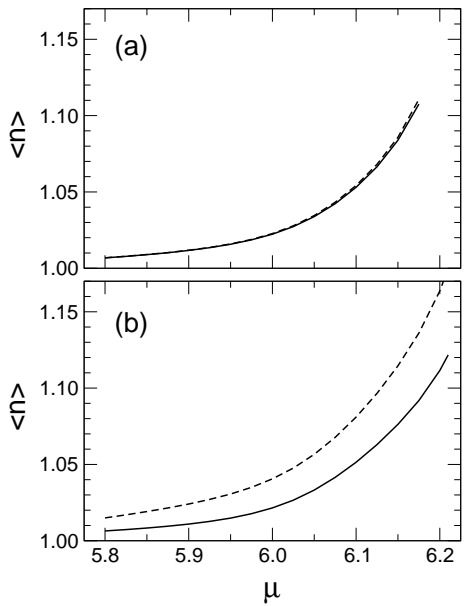}
\end{center}
\caption{\label{checkn} 
Filling $n = \langle n_{i\sigma} \rangle$ as a function of chemical potential $\mu$ 
obtained by integration of the spectral density (Eq.\ (\ref{eq:n2}), solid lines) and via 
the derivative of $\Omega$ (Eq.\ (\ref{eq:n1}), dashed lines). 
Results are obtained, respectively, by considering $\varepsilon'$ as a variational 
parameter (a), and by setting $\varepsilon'=0$ (b). 
Calculations for the electron-doped case with $U=8$, nearest-neighbor hopping $t_{nn}=-1$ 
and next-nearest-neighbor hopping $t_{nnn}=0.3$.
}
\end{figure}

The effect of thermodynamic (in)consistency is illustrated for the single-band 
Hubbard model of Eq.~(\ref{eq:hub}) in Fig.\ \ref{checkn}. 
The figure shows $n = \langle n_{i\sigma} \rangle$ as a function of $\mu$ as
obtained from VCA calculations described in Sec.\ \ref{sec:vca} below.
Solid lines display the result obtained by frequency integration of the spectral
density, Eq.\ (\ref{eq:n2}), while dashed lines show the result of the numerical 
$\mu$ derivative of $\Omega$, Eq.\ (\ref{eq:n1}).
We compare the results obtained by considering $\varepsilon'$ as a variational 
parameter (a), with the ones obtained by setting $\varepsilon'=0$ (b). 
In case (a) the two curves coincide within numerical accuracy, as expected, 
while in case (b) a considerable discrepancy is observed.
This discrepancy increases upon approaching the transition to the non-magnetic 
state, i.e.\ precisely in the interesting region, where it reaches about
$5\%
$ doping.

While thermodynamic consistency with respect to the particle number is an
issue for most approximations within the SFT, there is one exception: the cellular
DMFT. This is discussed in Appendix \ref{sec:cdmft}.

\subsection {Variational cluster approach (VCA)}
\label{sec:vca}

In detail, the calculation proceeds as follows: 
We consider the $2D$ single-band Hubbard model of Eq.~(\ref{eq:hub}) with 
nearest- ($t_{nn}$) and next-nearest ($t_{nnn}$) neighbor hoppings and 
the Hubbard $U$. 
For the purpose of our qualitative description of the HTSC, it is sufficient to
take parameters which are typical for both hole- and electron-doped high-$T_c$
cuprates, i.e.\ $t_{nnn}/t_{nn} = -0.3$ and $U/t_{nn}=8$. 
The energy scale is set by choosing $t_{nn}=-1$.
Different values of the parameters (e.g. a change of $U$ or $t_{nnn}$ within 
$\sim 30 \%
$) or a third-nearest neighbor hopping have been also incorporated, 
checked and found not to qualitatively change our conclusions (for example,
concerning the weak phase separation detected in the electron-doped case).

The Hamiltonian of the reference system $H'$ is given by a set of decoupled clusters
of finite size. 
For an individual cluster, the Hamiltonian reads:
\begin{equation}
  H'_{\rm cluster} = H'_{\rm Hub.} + H'_{\rm AF} + H'_{\rm SC} \: .
\label{eq:ham}
\end{equation}
It consists of the Hubbard Hamiltonian confined to the finite cluster $H'_{\rm Hub.}$ 
plus two symmetry-breaking terms (Weiss fields) $H'_{\rm AF}$ 
and $H'_{\rm SC}$ with
\begin{equation}
    H'_{\rm AF} = h'_{\rm AF} \sum_{i\sigma} (n_{i\uparrow} - n_{i\downarrow}) 
    e^{i \ff{Q} \ff{R_i}}
\label{haf}
\end{equation}
and
\begin{equation}
    H'_{\rm SC} = h'_{\rm SC} \sum_{ij} \frac{\eta_{ij}}{2} 
    (c_{i\uparrow} c_{j\downarrow} + \mbox{h.c.})
\label{hsc}
\end{equation}
where $h'_{\rm AF}$ is the strength of the staggered and $h'_{\rm SC}$ the strength
of the nearest-neighbor $d$-wave pairing field. 
$\ff{Q} = (\pi, \pi)$ is the AF wave vector, and $\eta_{ij}$ 
denotes the $d$-wave form factor which is non-vanishing for nearest-neighbor lattice 
sites only and is equal to $+1$ ($-1$) for $\ff R_i - \ff R_j$ in x (y) direction. 
The sum in Eq.\ (\ref{hsc}) is restricted to sites $i$ and $j$ belonging to the same 
cluster.
According to the discussion in Sec.\ \ref{sec:n}, the site-independent energy
$\varepsilon'=t'_{ii}$ shall be treated as a variational parameter.
It shows up in the local term 
\begin{equation}
    H'_{local} = \varepsilon' \sum_{i\sigma} n_{i\sigma}
\end{equation}
which is already included in $H'_{\rm Hub.}$.
The optimization of $\varepsilon'$ has to be done simultaneously with the optimization 
of the parameters $h'_{\rm AF}$ and $h'_{\rm SC}$.

Due to the optimization of the Weiss field strengths $h'_{\rm AF}$ and $h'_{\rm SC}$, 
one can account for spontaneous AF and $d$-wave SC symmetry breaking. 
The respective AF and SC order parameters, $m$ and $\Delta$, are defined as
\begin{equation}
  m = \frac{\partial \Omega }{ \partial h_{\rm AF} } \; , \qquad 
  \Delta = \frac{ \partial \Omega }{ \partial h_{\rm SC} } \; ,
\label{eq:op}
\end{equation}
in the limit $h_{\rm AF}, h_{\rm SC} \to 0$ where $h_{\rm AF}, h_{\rm SC}$ are the
strengths of external {\em physical} staggered and pairing fields.
These physical fields should not be confused with the ficticious Weiss fields with
strengths $h'_{\rm AF}$ and $h'_{\rm SC}$.
Adding the respective physical field terms to the Hamiltonian $H$ and performing the 
derivative with respect to $h_{\rm AF},h_{\rm SC}$ of the SFT grand potential (at the 
respective optimal ficticious field strengths $h'_{\rm AF},h'_{\rm SC}$), yields 
$m$ and $\Delta$ {\em consistently} with their representations (on the ``one-particle 
level'') as frequency integrals over the usual and anomalous one-particle spectral density.
This consistency is shown in the same way as in Sec.\ \ref{sec:n} for the average
particle number and is a consequence of treating the ficticious fields as variational
parameters.

The quality of the approximation is decisively influenced by the cluster size used.
On the one hand, for an appropriate characterization of the phase transition
(as considered in Sec.\ IV, below), one needs a sufficient accuracy in the grand 
potential $\Omega$.
This accuracy is, first of all, determined by the requirement that the clusters 
chosen must be large enough to fully account for the ``short-range'' spatial dependence 
of the self-energy as has been discussed below Eq.\ (\ref{eq:sf}).
On the other hand, for a given cluster size, an as accurate as possible numerical 
evaluation has to be employed. 

With respect to the latter, we found it both convenient and accurate to evaluate the
frequency integrals contained in the trace in Eq.\ (\ref{eq:ocalc}) not by
numerical integration (where the required accuracy is difficult to achieve), 
but by converting these integrals to a sum over the poles of the Green's function 
(see Ref.\ \onlinecite{pott.epj} for details). 
This, however, requires the computation of all many-body eigenstates in a given sector
of the cluster Hamiltonian $H'_{\rm cluster}$ (Eq.\ (\ref{eq:ham})) including
symmetry-breaking fields.
This technically limits the cluster sizes to be considered. 
Therefore, in the present work, we have chosen an infinite lattice tiled with 
$2\times 2$ clusters. 

Cluster consisting of $2\times 2$ sites, as well as larger cluster sizes have 
recently been systematically studied by Kyung et al.\ \cite{ky.ko.06} in their 
influence on various physical quantities of the $2D$ Hubbard model by means of 
the C-DMFT. 
Already the smallest, i.e.\ $2\times 2$, cluster has been found to account for 
more than 95\% of the ``correlation effect'' in the single-particle spectrum
(for a precise definition see Eq.~(16) of Ref.\ \onlinecite{ky.ko.06}). 
This suggests that at least some of the relevant questions in a strongly 
correlated electron system, modelled by a Hubbard-type Hamiltonian, may be studied, 
even rather accurately, with a such a small reference cluster. 

Nevertheless, for a more accurate determination of the SC state and of the phase
transition, larger clusters have to be considered eventually. 
Only in this way, phase-fluctuation effects between different d-wave Cooper pairs 
(which determine the spatial dependence of the self energy in the SC state) can be 
accounted for. 
We expect, for example, that calculations for larger clusters would display
a smaller SC gap than the one seen in Fig.~\ref{spectrelhol}b around $(\pi,0)$ 
in the electron-doped case. 
On the other hand, we expect our results on the global phase diagram to be quite
robust.
This is corroborated by the only weak dependence on (moderate) model changes,
such as the next-nearest hopping or the Hubbard interaction.\cite{se.la.05, ai.ar.05}

\section{Single-particle excitations}
\label{spe}
\begin{figure}[b]
\epsfig{file=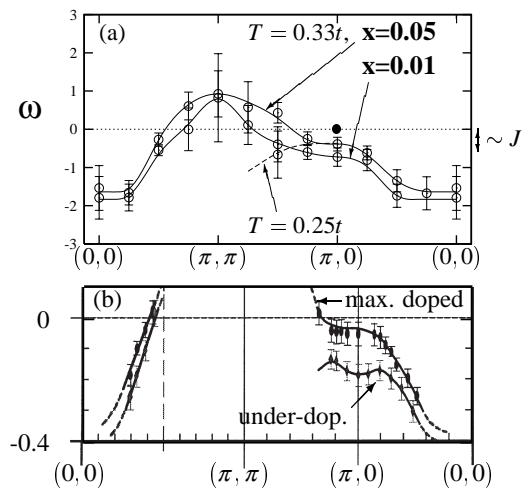,width=0.8\columnwidth}
\caption{Taken from Ref.\ \onlinecite{pr.ha.97}.
The dispersion of the peaks in the single-particle spectral
weight from (a) QMC simulations of the Hubbard model at the temperatures
indicated and at dopings $x=0.01$, $x=0.05$ (peaks in $A(\ff k,\omega)$ 
represented by error bars) and $x=0.13$ (solid circle).
(b) ARPES experiments from under-doped and optimally doped materials 
(peak centroids in $A(\ff k,\omega)$) after Ref.\ \onlinecite{ma.96}.}
\label{qmc}
\end{figure}
 
\begin{figure}[t]
\includegraphics[width=0.9\columnwidth]{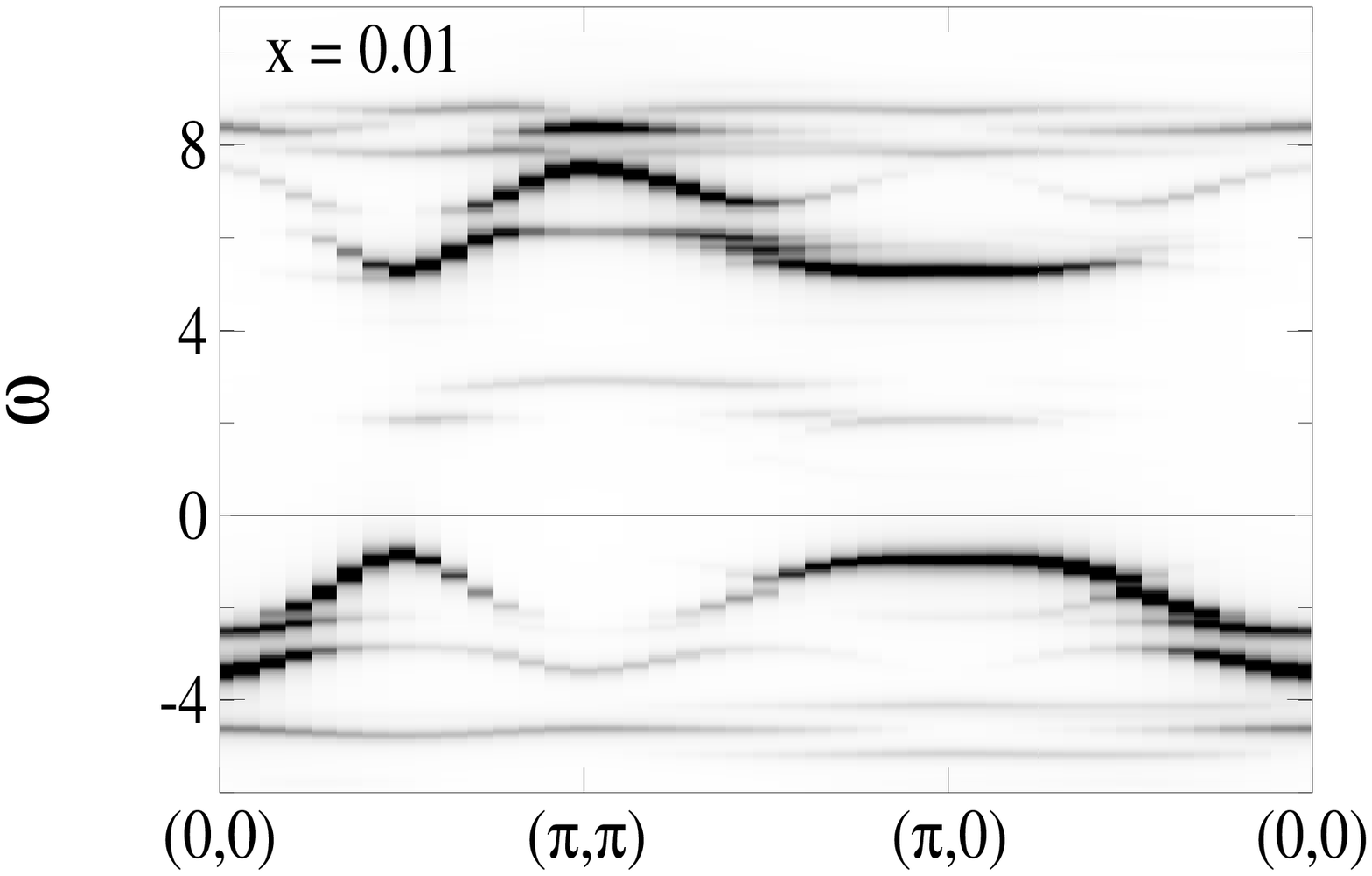}\\
\includegraphics[width=0.9\columnwidth]{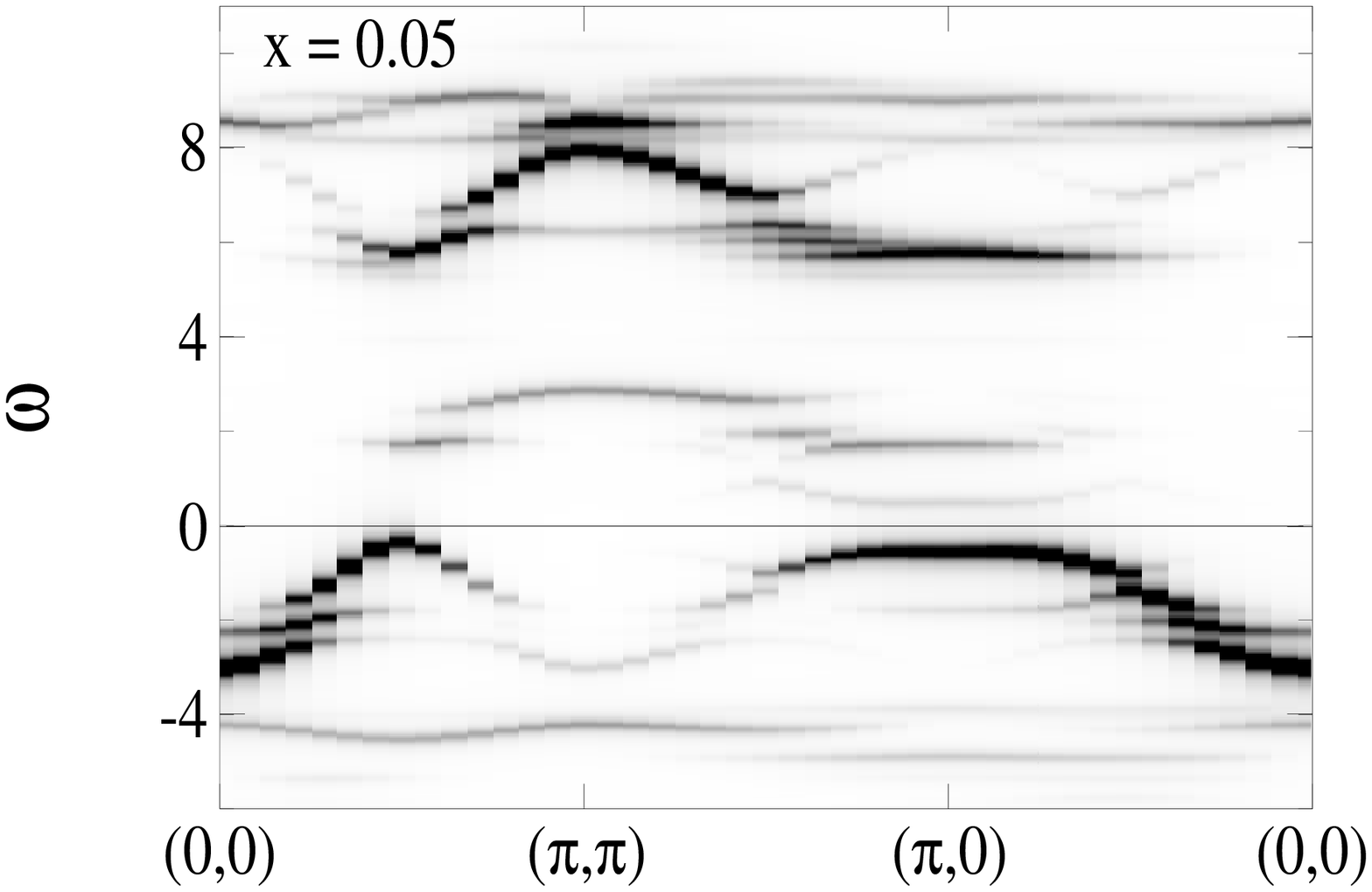}\\
\caption{
Single-particle spectrum for the hole-doped ($x = 0.01$ and $x=0.05$) case
at $U=8$ ($t_{nn}=-1$).
For comparison with the QMC data, the next-nearest-neighbor hopping 
is set to $t_{nnn}=0$.
}
\label{spectrtp0}
\end{figure}

The evolution of the single-particle excitation spectrum $A (\ff{k}, \omega)$
as a function of hole doping, and, in particular, the transition from a ``small'' 
Fermi surface (hole pockets around ($\pi/2,\pi/2$)) to an LDA-like Fermi surface 
closed around $(\pi,\pi)$ are key observations in unlocking the mystery of the 
cuprates.
Earlier QMC calculations for the $2D$ Hubbard model \cite{pr.ha.97,ande}
found that the single-particle spectral weight  
$A(\ff k,\omega)$ semi-quantitatively reproduces both the momentum
($d_{x^{2}-y^{2}}$-symmetry) and, in particular, doping dependence of
the ``high-energy'' pseudogap of the order of the exchange energy $J \sim 200 meV$
as found in photoemission experiments around ($\pi, 0$). 
The corresponding QMC data are reproduced in Fig.\ \ref{qmc} for comparison 
with the results of the VCA shown in Fig.\ \ref{spectrtp0} for hole dopings 
$x=0.01$ and $x=0.05$. 

We first discuss the QMC data.
In the under-doped regime, the ``pseudogap'' feature near ($\pi, 0$) moves to
lower binding energy as doping is increased. 
At about $x=0.13$ (full circle in Fig.\ \ref{qmc}) the pseudogap vanishes in 
overall accordance with the experimental findings. \cite{da.hu} 
In the experiments, the ``high-energy'' pseudogap is identified with the 
centroids of spectral weight near ($\pi, 0$). 

However, the experimental ARPES spectra also display a ``low-energy'' pseudogap 
in the normal (superconducting) state above (below) $T_c$ with energy $\sim 20 meV$, 
inferred from the leading edge in the spectral density, which also opens up in 
the under-doped regime and vanishes in the over-doped regime. \cite{da.hu} 
This empirical correlation between the disappearance of the ``high-energy'' 
pseudogap and the decrease of the SC gap and, therefore, pairing strength 
suggested already several years ago that the high-energy features at
($\pi, 0$) (which are correctly described by the ``high-temperature''
QMC simulations) are closely related to the pairing interaction. \cite{da.hu}

It is clear, however, that there is a strong need to perform calculations 
at much lower temperatures, i.e.\ temperatures below $T_c \sim 20 meV$.
Only then the ``low-energy'' SC or normal state pseudogap can be detected 
and only then the question ``where do holes enter first?'' can be answered 
correctly.
According to the results of the ``high-temperature'' ($T = 0.33$) QMC simulations 
shown in Fig.\ \ref{qmc} for $x=0.01$, holes enter first around ($\pi, \pi$).
This can be understood by referring to the idea \cite{pr.ha.95,pr.ha.97} that a higher 
temperature $T$ effectively acts as an increased doping which destroys the 
magnetic Brillouin zone. 
This allows holes to first enter into the ``arc'' of single-particle
excitations spanned around $\ff{Q} = (\pi, \pi)$.

\begin{figure}[t]
\includegraphics[width=0.9\columnwidth]{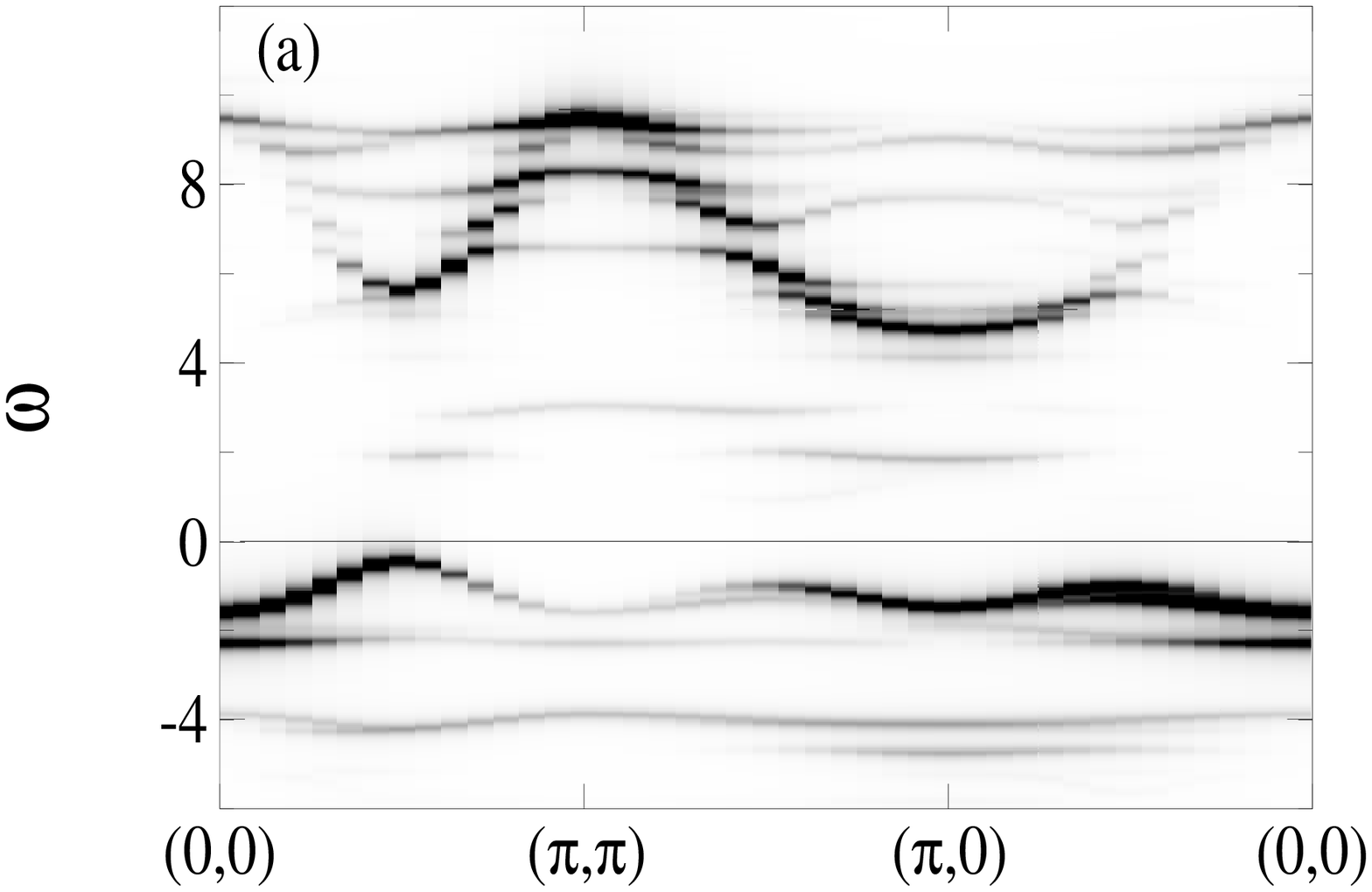}
\includegraphics[width=0.9\columnwidth]{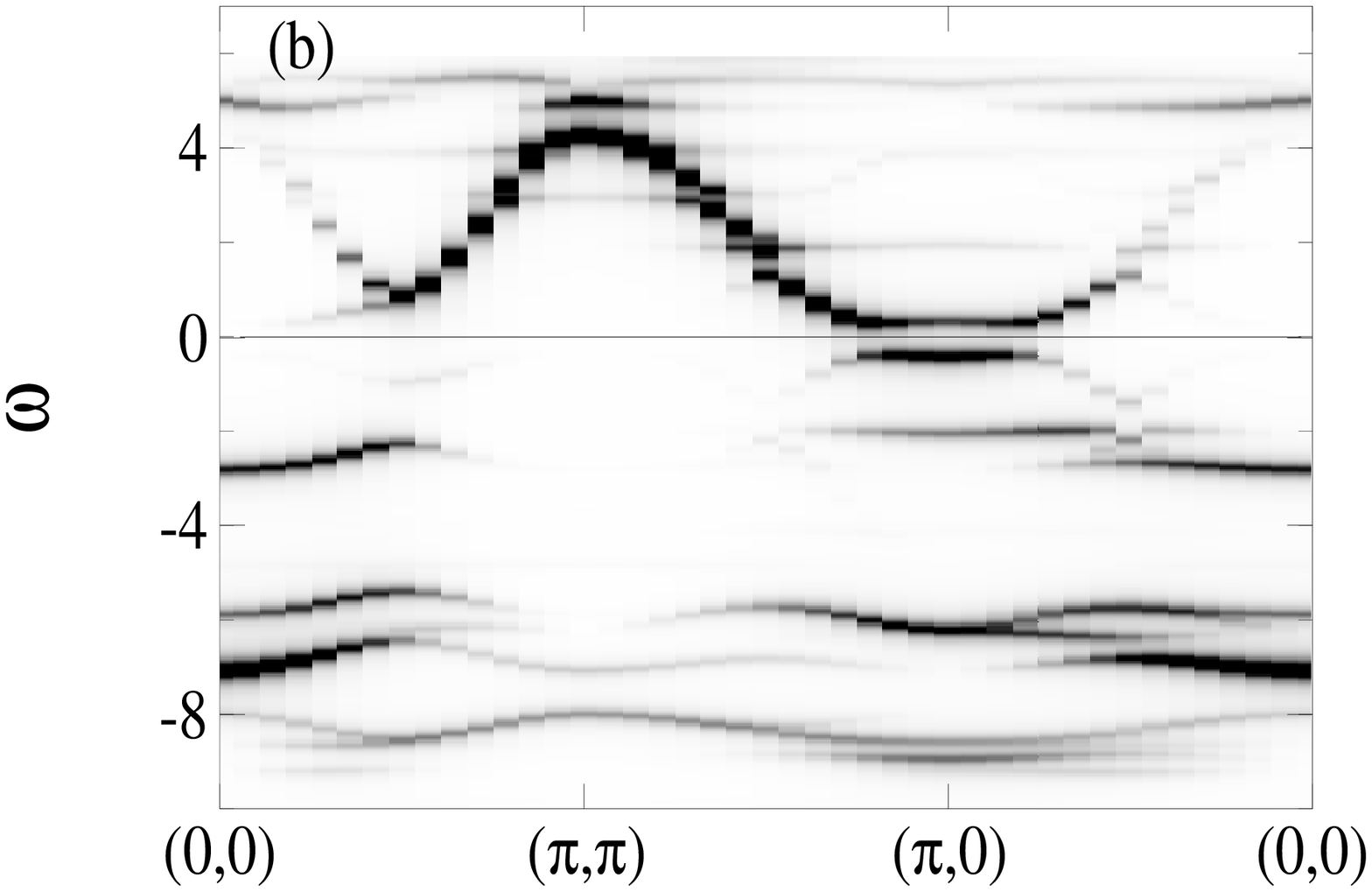}
\caption{
Single-particle spectrum for $U=8$ and $t_{nnn}=0.3$ ($t_{nn}=-1$)
in the hole-doped (a) and in the electron-doped (b) case.
Results are shown for dopings in the mixed AF+SC state (see Sec.\ \ref{gspd}), i.e.\ for 
$x = 0.015$ (a) and $x=0.09$ (b), respectively.
}
\label{spectrelhol}
\end{figure}

However, a portion of the corresponding ``large'' Fermi surface seems to disappear 
already at the lower temperature $T=0.25$. 
This is indicated in Fig.\ \ref{qmc}(a) by the downturn of the quasi-particle-like 
band between ($\pi, 0$) and ($\pi, \pi$). In the experiment (Fig.\ \ref{qmc}(b)), 
this behavior in the under-doped regime has been interpreted as the opening of a 
pseudogap in the underlying Fermi surface near the ($\pi,0$) to ($\pi, \pi$) line. 
\cite{da.hu}
This opening is obviously intimately related to the above question,
namely where doped holes first enter, i.e.\ to the possibility of hole pockets
developing at very small temperatures and low dopings around ($\pi/2,\pi/2$) in the 
hole-doped case.

In contrast to the finite-$T$ QMC result, the corresponding VCA calculations for $T=0$
(see Fig.\ \ref{spectrtp0}) show that holes first go into the coherent band around 
($\pi/2, \pi/2$) forming ``hole pockets'' consistent with experiments. \cite{da.hu} 
Note that for the dopings considered in Fig.\ \ref{spectrtp0}, the system is in a 
mixed AF+SC state with non-vanishing AF order parameter (see also Fig.~\ref{phas-hole}). 
Therefore, although the low-lying spectral weight is found around $(\pi/2,\pi/2)$, 
the spectrum still shows an AF gap at this wave vector. 

Related to this, it appears that even for $x=0.05$ the quasi-particle-like dispersion 
does not cross the Fermi energy along the nodal direction (between ($0,0$) and ($\pi,\pi)$) 
in contrast to the experimental situation in Fig.\ \ref{qmc}b. 
Although some (weak) accumulation of spectral weight close and above $\omega=0$ can be 
seen in Fig.\ \ref{spectrtp0}, we believe that this non-crossing is (partly) a shortcoming 
of our $2 \times 2$ calculation and that the self-energy in the nodal
direction requires larger clusters. 

To address this issue for more realistic model parameters, 
i.e.\ including a next-nearest-neighbor hopping $t_{nnn}=-0.3 t_{nn}$, as well as 
to study the corresponding doping evolution in the electron-doped case, we discuss the 
VCA results for $A (\ff{k}, \omega)$ shown in Fig.\ \ref{spectrelhol}.
The calculations have been performed for the hole- (a) and for the electron-doped 
system (b) in the mixed AF+SC phase (see Sec.\ \ref{gspd}), i.e.\ for $x=0.015$ 
and $x=0.09$, respectively.

Let us concentrate on the very small ($1.5 \%
$) hole doping first.
In agreement with the corresponding experiments \cite{da.hu} in hole-doped 
cuprates, holes indeed first enter at ($\pi/2,\pi/2$).
Although the system is in a mixed AF+SC phase (see Sec.\ \ref{gspd}), 
the SC gap is zero at this nodal point, so that doping into nodal states,
i.e.\ normal metallic screening, apparently destabilizes the AF
solution already for rather low doping values ($x \sim 0.03$). 
This explains the phase diagram of Fig.\ \ref{phas-hole}.

\begin{figure}[t]
\begin{center}
        \includegraphics[width=0.98\columnwidth]{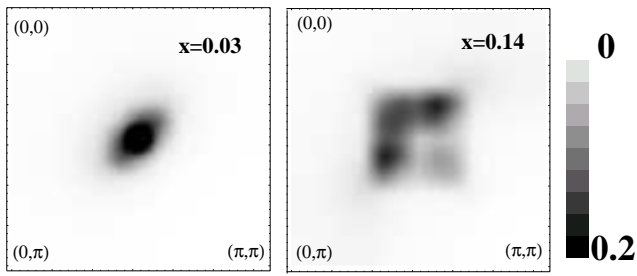}
\end{center}
\caption{\label{fermis-holdop} 
Evolution of the low-energy spectrum upon hole doping.
Parameters are as in Fig.\ \ref{spectrelhol}. 
The weight is obtained by integrating the low-energy $A (\ff{k},
	\omega)$-spectrum down to $0.2$ below the Fermi energy.
}
\end{figure}

\begin{figure}[b]
\begin{center}
        \includegraphics[width=0.98\columnwidth]{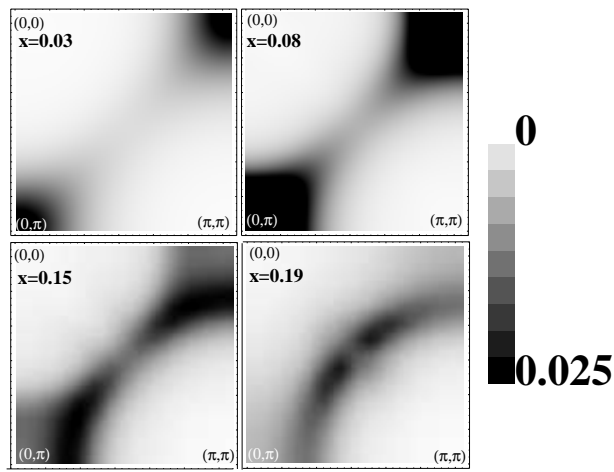}
\end{center}
\caption{\label{fermis-eldop} 
The same as in Fig.\ \ref{fermis-holdop} but for the electron-doped case.
}
\end{figure}

A similar picture can be inferred by looking at the evolution of the
Fermi surface as a function of doping. 
This can be extracted from Fig.\ \ref{fermis-holdop} where the low-energy 
spectral weight is plotted in the Brillouin zone.
In qualitative agreement with experiments, hole pockets start forming
around $(\pi/2,\pi/2)$ for low doping, while a large Fermi surface
centered about $(\pi,\pi)$ starts building up at higher doping.

In contrast, in electron-doped systems, doped electrons initially form pockets
around ($\pi, 0$) (see Figs.\ \ref{spectrelhol}(b) and \ref{fermis-eldop}, 
see also Refs.\ \onlinecite{ku.ma.02,se.la.05}), in agreement with experiments.
\cite{ar.ro.02,ar.lu.01} 
Fig.\ \ref{fermis-eldop} shows the spectral weight obtained by integrating 
$A (\ff k,\omega)$ down to $0.2$ below the Fermi energy ($\omega=0$).
One has to be careful when interpreting Fig.\ \ref{fermis-eldop}, since for $x<0.13$ 
there is still an AF gap at the Fermi surface near $(\pi/2,\pi/2)$ 
besides the SC one near $(\pi,0)$.
Therefore, the excitation spectrum is completely gapped, and one has to go away 
from $\omega=0$ in order to find some weight.
Since we integrate only in a small energy window around the Fermi level, the
scale in Fig.\ \ref{fermis-eldop} is an order of magnitude smaller than in the 
hole-doped case, Fig.\ \ref{fermis-holdop}, for which the AF gap near
$(\pi/2,\pi/2)$ is shifted away from the Fermi surface.
Nevertheless, one finds
that the lowest-lying states in the electron-doped case are around $(\pi,0)$.
Here, the density of states is large and provides a large ``reservoir'' for electron
doping. 
This, in combination with the fact that the chemical potential lies in the SC gap 
(of the AF+SC phase), stabilizes the AF solution for a larger doping range than in 
the hole-doped case, allowing for the AF gap to decrease more gradually 
(see Sec.\ \ref{gspd}).
Also in this case, the breakdown of the magnetic solution occurs as soon as 
the chemical potential reaches the bottom of the band at ($\pi/2,\pi/2$).
The observation that doping into ($\pi/2,\pi/2$) generically makes the AF 
phase unstable, suggests that also the occurrence of phase separation, to be
discussed in Sec.\ \ref{gspd}, is rather generic and quite independent of
model details.

The low-energy spectral weight, as plotted for different dopings in
Fig.\ \ref{fermis-eldop}, agrees quite well with the ARPES measurements of 
Armitage et al.\ \cite{ar.ro.02,ar.lu.01};
doped electrons away from half-filling first form pockets around $(\pi,0)$. 
At higher doping, Fermi-surface segments start developing along the 
$(\pi,0)-(0,\pi)$ line.
Finally, these segments connect and form a large Fermi surface around
$(\pi,\pi)$ near optimal doping.

\section{Ground-state phase diagram}
\label{gspd}

In this section the ground-state phase diagram and the AF to SC transition are
presented and discussed (see also Refs.\ \onlinecite{se.la.05,ai.ar.05}). 
In particular, we will focus on the similarities and differences between the 
hole- and electron-doped system as well as their relation with the evolution 
of the single-particle spectrum.

In order to determine the $T=0$ phase diagram, we proceed as described in 
Sec.\ \ref{sec:vca}, i.e.\ two symmetry-breaking terms (Weiss fields) 
$H'_{\rm AF}$ and $H'_{\rm SC}$ are included and their respective strengths,
$h'_{\rm AF}$ and $h'_{\rm SC}$ are treated as variational parameters
in addition to $\varepsilon'$.
As discussed in Sec.~\ref{sec:n}, the use of the additional variational 
parameter $\varepsilon'$ is required in order to have a consistent determination
of the particle density. 

\begin{figure}[t]
\begin{center}
\includegraphics[width=0.8\columnwidth]{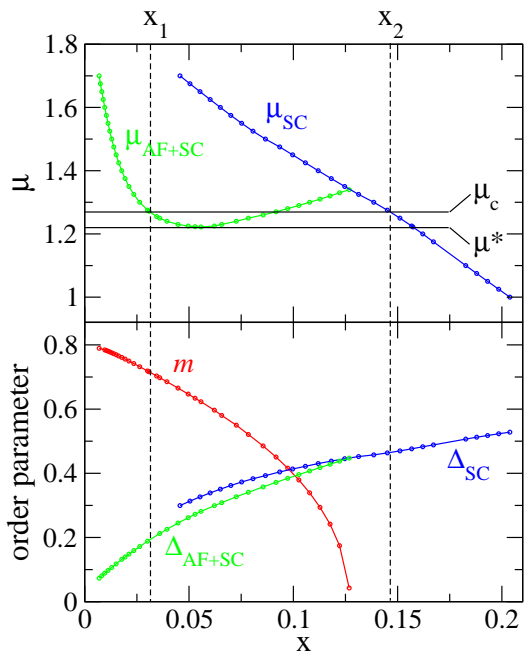}
\end{center}
\caption{\label{phas-hole} 
Antiferromagnetic and superconducting order parameters, $m$ and $\Delta$,
and chemical potential $\mu$ as functions of hole doping $x$. 
$\Delta$ and $\mu$ are plotted for the AF+SC 
(green, $\Delta_{\rm AF+SC}$, $\mu_{\rm AF+SC}$) as well
as for the pure SC homogeneous solutions (blue, $\Delta_{\rm SC}$, $\mu_{\rm SC}$). 
Note that $\Delta$ is scaled by a factor 5 for convenience. 
For $x<x_1$ the system exhibits a coexistence of AF and d-wave SC order.
Phase separation occurs between the doping levels $x_1$ and $x_2$.
For $x>x_2$ pure d-wave SC is realized.
In the phase separation region $x_1 < x < x_2$, the homogeneous solution 
becomes unstable, and the system prefers to separate into a mixture of two 
densities corresponding to $x_1$ and $x_2$. 
The chemical potential $\mu_c$ is determined by the Maxwell construction 
shown in the upper figure. 
At $\mu^\ast$ the slope of the AF+SC solution changes sign.
}
\end{figure}

\begin{figure}[t]
\begin{center}
\includegraphics[width=0.8\columnwidth]{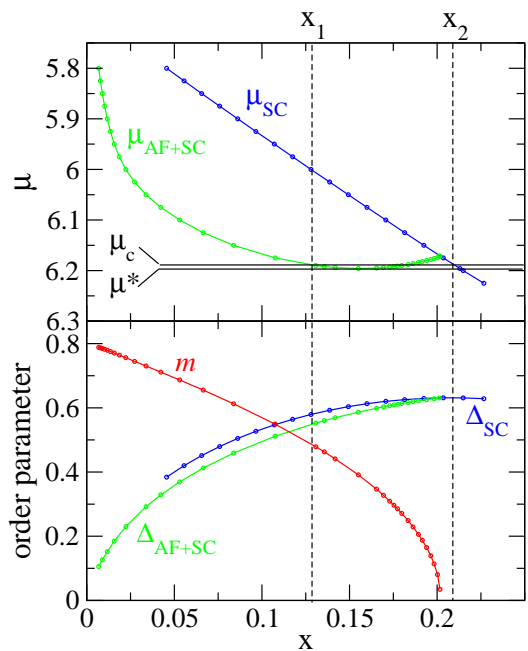}
\end{center}
\caption{\label{phas-elec} 
Same as Fig.~\ref{phas-hole} but for electron doping.
Note the enhanced robustness of the AF state and the 
strongly reduced scale 
$\Delta \mu \equiv (\mu^\ast - \mu_c)$ as compared to hole doping.
}
\end{figure}

The phase diagram for the Hubbard model with $U=8$ and next-nearest-neighbor hopping
$t_{nnn}=0.3$ (we set $t_{nn}=-1$) as obtained from our calculations, is shown 
in Fig.\ \ref{phas-hole} for the hole-doped and in Fig.\
\ref{phas-elec} for the 
electron-doped case.
In the upper part of each figure, the chemical potential $\mu$ is plotted as a 
function of $x$. 
In the corresponding lower parts, we display the AF $(m)$ and SC $(\Delta)$ order 
parameter as a function of doping $x$.
Note that the order parameters $m$ and $\Delta$ are different from the 
Weiss fields $h'_{\rm AF}$ and $h'_{\rm SC}$, respectively.
Quite generally, however, a nonvanishing stationary value for a Weiss field
produces a nonvanishing order parameter, respectively, although the latter 
can be much smaller.

Let us discuss hole doping first (see Fig.\ \ref{phas-hole}). 
For low dopings $x$ we find a homogeneous symmetry-broken phase in
which both the AF as well as the SC order parameter $m$ and $\Delta$ are non-zero.
This corresponds to a phase where AF and SC order microscopically and coherently 
coexist.
A homogeneous phase with pure SC ($m=0$ and $\Delta >0$) is obtained for larger dopings. 
The behavior of $m$ vs.\ $x$ in Fig.\ \ref{phas-hole} seems to suggest that the transition 
to the non-magnetic state is continuous (second order) as a function of doping.
However, a glance at the non-monotonous behavior of the chemical potential $\mu$
plotted as a function of $x$ in the upper part of the figure, indicates the occurrence 
of a charge instability.
The system tends to separate into a hole-poor ($x_1$) and a hole-rich ($x_2$) phase.
The two dopings $x_1$ and $x_2$, as well as the chemical potential $\mu_c$ in the 
phase-separated region, are identified by the Maxwell construction shown
in the upper 
part of Fig.\ \ref{phas-hole}.
$\mu^\ast$ is the point where the slope of $\mu(x)$ changes sign.

Of course, this phase separation is obtained within a treatment that,
in the present work, just considers for homogeneous order parameters
only and that neglects surface effects.
Furthermore, the VCA is a mean-field-type approach on length scales beyond the 
size of the individual cluster.
Therefore, the above result has to be interpreted with care. 
We expect it to signal a {\it tendency} towards the formation of microscopic 
inhomogeneities, such as stripes, checkerboard patterns, 
etc.\ \cite{ca.em.03,ar.ki.03,ar.fr.04} This tendency will be further
studied in future work by considering larger clusters and/or by
allowing for a more general variational solution which explicitly describes 
stripe inhomogeneities
(for example, by considering coupled clusters with different dopings,
as in Ref.\ \onlinecite{za.ed.00}). In this case, one might expect the phase
transition to become more continuous and phase separation to disappear eventually.

Let us now discuss the electron-doped case and the similarities and differences 
as compared to hole-doping.
The first observation is that, while the phase diagrams in Figs.\ \ref{phas-hole} 
and \ref{phas-elec} are {\em qualitatively} similar, the phase in which long-range 
AF order is realized is spreading to significantly larger doping values in the 
electron-doped case, in overall agreement with the experimental situation.
Fig.\ \ref{phas-elec} shows that phase separation
occurs in the electron-doped case as well, although the associated energy scale 
$\Delta \mu \equiv (\mu^\ast - \mu_c)$ is smaller with respect to the hole-doped 
case by about an order of magnitude.
In Ref.\ \onlinecite{ai.ar.05} it is argued that this can be related to the different 
pseudogap and SC transition scales in hole- and electron-doped materials. 
This may give support to theories \cite{in.do.88,ca.em.03} which are based on the notion
that fluctuations of competing phases, or of the related order parameters, are responsible 
for the pseudogap phenomenon.

\begin{figure}[t]
  \includegraphics[width=0.7\columnwidth]{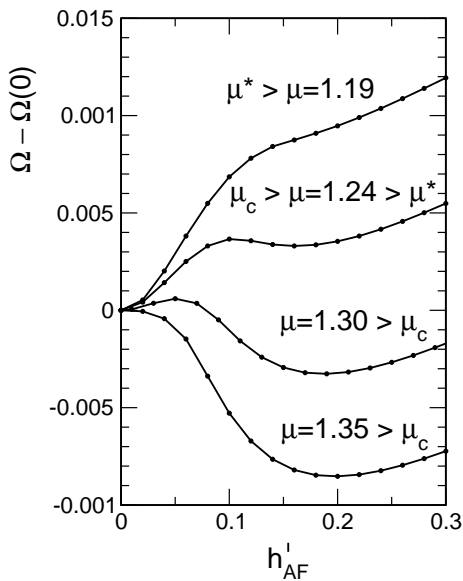}
\caption{
$\Omega$ vs.\ $h'_{\rm AF}$ 
(with the two other parameters $h'_{\rm SC}$ and $\varepsilon'$ fixed at 
their stationary-point values)
for different $\mu$ in the hole-doped case.
Parameters: $U=8$, $t_{nnn}=0.3$ ($t_{nn}=-1$).
}
\label{omegamin}
\end{figure}

Depending on the value of $\mu$, there may 
be two solutions, an AF+SC and a pure SC one, corresponding to two stationary points 
of $\Omega$.
In order to gain insight into the first-order transition between these two solutions, 
it is instructive to observe the behavior of the grand potential $\Omega$ as a function 
of the variational parameter $h'_{\rm AF}$ associated with this transition for 
different $\mu$.
For simplicity, we consider here the hole-doped case only.
Results are plotted in Fig.\ \ref{omegamin}. 
Note that $h'_{\rm AF}$ is varied, while the other two variational parameters,
$h'_{\rm SC}$ and $\varepsilon'$, are fixed at their values at the respective
stationary point.
At low doping, there is a single minimum of the grand potential 
$\Omega$ at a finite value $h'_{\rm AF}$ only (Fig.\ \ref{omegamin}, $\mu=1.35$).
In contrast, the paramagnetic ($h'_{\rm AF}=0$) solution is given by a local maximum.
Upon further doping, $\Omega$ \emph{additionally} develops a local minimum at 
vanishing AF variational parameter $h'_{\rm AF}=0$ (Fig.\ \ref{omegamin}, $\mu =1.30$).
For $\mu < \mu_c$, the minimum at $h_{\rm AF}=0$ becomes lower than the one at finite 
$h'_{\rm AF}$ indicating a first-order phase transition to a non-magnetic state 
(Fig.\ \ref{omegamin}, $\mu =1.24$). 
Eventually, the local maximum lying between the two minima merges with the minimum
at finite $h'_{\rm AF}$ and \emph{the AF solution disappears}
(Fig.\ \ref{omegamin}, $\mu =1.19$).
This disappearance just occurs when the chemical potential $\mu$ enters the
quasi-particle band around $(\pi/2,\pi/2)$ ($\mu^\ast = 1.22$).

Although Fig.\ \ref{omegamin} displays the results for the hole-doped case, 
this behavior occurs qualitatively in the electron-doped case as well, 
although with a much smaller characteristic energy scale as discussed above.
Again, these results reflect a qualitatively similar behavior for electron and
hole doping.

\section{Summary and Conclusions}
\label{summ}

By means of a recently developed quantum-cluster approach, we have carried out a 
detailed analysis of the phase transition from the antiferromagnetic to the 
superconducting phase in the Hubbard model at zero temperature.
The main results concerning the nature of the AF to SC transition are summarized 
in Figs.\ \ref{phas-hole} and \ref{phas-elec}.
At low dopings the AF phase actually mixes with a weak $d$-wave SC component. 
A similar coexistence phase is observed, for example, in PrCeCuO, \cite{da.ka.05} 
and was also obtained in previous mean-field and cluster calculations. 
\cite{li.ka.00,se.la.05, in.do.88}
Upon further increasing doping, we find a transition to a pure $d$-wave SC phase. 

The phase-separation scenario, which is found for hole but also for electron doping 
within our approximation, should carefully be interpreted as a general
\emph{tendency} of the 
system to form microscopically inhomogeneous phases.
Indeed, by allowing for a more general spatial dependence of the order parameter, 
the macroscopic phase separation will probably be replaced by other microscopically 
inhomogeneous phases, such as stripes, checkerboard order, etc. 
In particular, this might be expected to be the case if long-range Coulomb interaction 
is taken into account additionally. \cite{ca.em.03,lo.em.94,ar.ha.02} 
The situation is subtle in the electron-doped case. 
Here, in contrast to previous theoretical calculations, our results also suggest 
phase separation, although the corresponding energy scale (see Fig.\ \ref{phas-elec}) 
is one order of magnitude smaller than in hole-doped compounds.

Our results for the single-particle excitations presented in Sec.\ \ref{spe} support 
the idea that it is, in fact, the single-particle spectrum which holds the key for
understanding the qualitative differences seen in the ground-state phase diagram 
between the hole- and electron-doped cuprates.
This concerns, for example, the robustness of the AF state in the electron-doped case.  
Here, electrons initially form pockets around ($\pi, 0$) in accordance with experiments.
\cite{ar.ro.02} 
The fact that the density of states is large there, as well as the presence of a 
SC ``gap'' at this nodal point, stabilizes the AF state for a larger doping regime 
as compared to the hole-doped case.
The breakdown of the magnetic solution appears as soon as ``normal'' 
metallic 
screening sets in, i.e.\ when $\mu$ touches the band at ($\pi/2, \pi/2$). 
\cite{ai.ar.05}

One should note that the variational cluster approach (VCA) is able to treat 
the fluctuations correctly up to the range of the cluster size only.
Therefore, the question arises whether the SC solution we (and also others 
\cite{ch.jo.90,li.ka.00,se.la.05}) obtain within the AF phase is a true 
long-range SC phase or whether it is only a signal of strong pairing fluctuations 
within the AF phase leading to a SC pseudogap.
The latter hypothesis could be supported by the fact that results obtained
with different cluster sizes \cite{se.la.05} seem to indicate a size
dependence of the SC order parameter, and by the fact that the SC order
parameter is about a factor three smaller in the AF+SC phase than in the pure
SC one. 
The presence or not of such a microscopic coexistence phase may depend
on material details. 
Certainly, our results suggest that the SC gap $\Delta$ (or pseudogap in the case of 
fluctuations) is important in order to stabilize the AF phase in 
electron-doped materials, as discussed in Sec.\ \ref{spe}.

Similar VCA calculations have recently been carried out by S\'{e}n\'{e}chal et al.\ 
\cite{se.la.05} using clusters up to 10 sites but without the variation of the 
on-site energies.
These authors show that the single-band Hubbard model is sufficient to explain the 
different overall shapes of the phase diagrams for hole- and electron-doped cuprates. 
However, their results seem to suggest that in the electron-doped case the AF to SC 
transition is continuous and associated with a quantum-critical point, in contrast 
to our results.
This clearly shows the importance for a consistent determination of the average 
particle number.

The recent substantial progress in relating the ``high-energy'' physics of the 
Hubbard model and its variants to the low-energy physics of competing phases is 
to a large extent due to the development of different quantum-cluster theories.
Apart from the VCA, these are the cluster extensions of the DMFT, such as the 
cellular DMFT (C-DMFT) and the dynamical cluster approximation (DCA). Kyung et al.\
\cite{ky.ko.06} have shown that some of the important problems in strongly 
correlated electron systems may be studied highly accurately using comparatively small
clusters.
Maier et al.\ \cite{ma.ja.05u,ma.ja.05u.ps} performed a systematic cluster-size 
study of the $2D$ Hubbard model using rather large clusters (up to $26$ sites). 
Converged results suggest a finite-$T$ instability to $d$-wave SC state. 
Because of the QMC minus-sign problem, however, results were limited
to $U=4t$ where the typical correlation energy $U$ and the magnetic
energy scale of the HTSC is not yet achieved. 
On the other hand, the present VCA studies are clearly not yet
converged with respect to the cluster size, as one can read off from,
for example, the relatively large SC gap in the single-particle
excitations, displayd in Fig.~\ref{fermis-holdop}. An extension to
larger clusters is, at least in principle, also possible within the VCA.
This, however, necessarily implies the use of stochastic (QMC) methods
as solvers for the cluster reference system.

\acknowledgments

We would like to thank D.\ J.\ Scalapino for many useful discussions.
One of us (WH) would additionally like to acknowledge the hospitality of the Kavli
Institute for Theoretical Physics in Santa Barbara (supported by NSF 
Grant No.\ PHY99-0794).
The work was supported by the DFG Forschergruppe 538, by the KONWIHR
supercomputing network in Bavaria and by the FWF Project N.\ P18505-N16.

\appendix

\section{One-particle density matrix}
\label{sec:cicj}

The reasoning in Sec.\ \ref{sec:n} can straightforwardly be generalized 
to the off-diagonal ($i\ne j$) elements of the one-particle density matrix 
$\langle c_{i\sigma}^\dagger c_{j\sigma} \rangle$.
In this case, thermodynamic consistency means that, for a selected pair
of sites $(i,j)$, the derivative
\begin{equation}
  \langle c_{i\sigma}^\dagger c_{j\sigma}\rangle
  = 
  \frac{\partial \Omega}{\partial t_{ij\sigma}} 
\label{eq:hop1}
\end{equation}
is equivalent with the integral
\begin{equation}
  \langle c_{i\sigma}^\dagger c_{j\sigma}\rangle
  = 
  \int_{-\infty}^\infty f(\omega) \: A_{ji\sigma}(\omega) \: d\omega \: ,
\label{eq:hop2}
\end{equation}
where, for convenience, the hopping is formally assumed to be spin-dependent.

Starting with Eq.\ (\ref{eq:hop1}), we note that the $t_{ij\sigma}$ dependence
of $\Omega \equiv \Omega_{\ff t, \ff U} [\ff \Sigma_{\ff t'_{\rm s},\ff U}]$ is due to 
the explicit $t_{ij\sigma}$ dependence of the free Green's function $\ff G_{0,\ff t}$
in Eq.\ (\ref{eq:ocalc}), and due to the implicit $t_{ij\sigma}$ dependence 
of $\ff t'_{\rm s}$:
\begin{eqnarray}
  \langle c_{i\sigma}^\dagger c_{j\sigma}\rangle
  & = &
  \frac{\partial \Omega_{\ff t, \ff U} [\ff \Sigma_{\ff t'_{\rm s},\ff U}]}{\partial t_{ij\sigma}} 
  \nonumber \\
  & = &
  T \sum_{\omega_n} e^{i\omega_n 0^+} \left(
  \frac{1}{\ff G^{-1}_{0,\ff t}(i\omega_n) - \ff \Sigma_{\ff t'_{\rm s}, \ff U}(i\omega_n)} 
  \right)_{ji\sigma}
  \nonumber \\
  & + &
  \frac{\partial \Omega_{\ff t, \ff U} [\ff \Sigma_{\ff t',\ff U}]}
  {\partial \ff t'} \Bigg|_{\ff t' = \ff t'_{\rm s}} \cdot
  \frac{\partial \ff t'_{\rm s}}{\partial t_{ij\sigma}} \: .
\end{eqnarray}
Because of the stationarity of the grand potential, 
the second term on the r.h.s.\ can be ignored.
We are thus left with the first term which, after transforming the Matsubara sum
into an integration over real frequencies, exactly yields the desired expression 
(\ref{eq:hop2}).

In case that not all elements of $\ff t'$ are treated as variational parameters,
the derivative of $\Omega$ in the second term on the r.h.s.\ is a derivative 
with respect to those elements only. 
If, for the given pair of sites $i$ and $j$, $t_{ij\sigma}'$ is selected as a
variational parameter, the contribution due to the implicit $t_{ij\sigma}$ dependence  
vanishes again. 
If, on the other hand, $t_{ij\sigma}'$ is not treated as a variational parameter 
but fixed at $t'_{ij\sigma}=t_{ij\sigma}$ from the beginning, the second term will 
give a finite contribution which spoils the thermodynamic consistency.

\section{Cellular DMFT}
\label{sec:cdmft}

The C-DMFT is obtained by choosing as a reference system disconnected clusters 
with a continuous non-interacting bath coupled to each cluster site.
Carrying out the $\ff t'$ partial derivatives in Eq.\ (\ref{eq:stat}) and using
Eq.\ (\ref{eq:ocalc}) for the SFT grand potential, we get the Euler equation in 
the form:
\begin{equation}
   T \sum_{\omega_n} \sum_{ij\sigma}
   \left( 
   \frac{1}{{\bm G}_{0, \ff t}^{-1} - {\ff \Sigma}_{\ff t',\ff U}}
   -  {\bm G}_{\ff t',\ff U} \right)_{ji\sigma} 
   \frac{\partial \Sigma_{ij\sigma}} 
        {\partial {{\bm t}'}}
   = 0 \: .
\label{eq:euler}
\end{equation}
Here $i$ and $j$ run over the sites of the original lattice, excluding bath
sites.
The one-particle bath parameters, namely the on-site energies of the bath sites
and the hybridization of the bath sites with the correlated (physical) sites,
have to be treated as (a continuous set of) variational parameters. 
In the C-DMFT it is assumed that bath parameters can be found such that the 
first factor in (\ref{eq:euler}) vanishes: 
\begin{equation}
   \Big( \frac{1}{{\bm G}_{0, \ff t}^{-1} - {\ff \Sigma}_{\ff t',\ff U}}
   \Big)_{ij\sigma} (\omega)
   =  
   \Big( {\bm G}_{\ff t',\ff U} \Big)_{ij\sigma} (\omega)
\label{eq:cdmft}
\end{equation}
for arbitrary $\omega$ and 
for sites $i$ and $j$ belonging to the same cluster.
Note that, by construction of the reference system, 
$\Sigma_{ij\sigma}(\omega)$ and also 
$\partial \Sigma_{ij\sigma}(\omega) / \partial {\bm t}'$ vanish
if $i$ and $j$ belong to different clusters. 
Consequently, if bath parameters can be found such that Eq.\ (\ref{eq:cdmft}) holds,
the Euler equation (\ref{eq:euler}) will be satisfied, too:
The self-energy functional is stationary at the C-DMFT self-energy.
Eq.\ (\ref{eq:cdmft}) is just the self-consistency equation of the C-DMFT
(see Refs.\ \onlinecite{ko.sa,po.ai.03}).

It is easy to see that $\langle N \rangle = \langle N \rangle'$ within the
C-DMFT:
This simply follows by comparing Eqs.\ (\ref{eq:nn}) and (\ref{eq:nnp}) for 
$\langle N \rangle$ and $\langle N \rangle'$ and by using the C-DMFT
self-consistency equation (\ref{eq:cdmft}).

Consider now the high-frequency expansions of the Green's function and of the
self-energy of the reference system. 
Using $\ff G' \equiv \ff G_{\ff t',\ff U}$,
$\ff \Sigma \equiv \ff \Sigma_{\ff t',\ff U}$,
and $\ff G_0 \equiv \ff G_{0,\ff t}$ for short,
we have (see Ref.\ \onlinecite{po.he} for high-frequency expansions in the Hubbard
model):
\begin{eqnarray}
G'_{ij\sigma}(\omega) &=& \frac{\delta_{ij}}{\omega} 
+ \frac{t'_{ij} - \mu \delta_{ij} + U \langle n_{i-\sigma} \rangle' \delta_{ij}}{\omega^2} 
+ {\cal O}(\omega^{-3}) \; ,
\nonumber \\
\Sigma_{ij\sigma}(\omega) &=& U \langle n_{i-\sigma} \rangle' \delta_{ij} + {\cal O}(\omega^{-1})\; .
\end{eqnarray}
Here, $\langle n_{i-\sigma} \rangle'$ is the average occupation in the reference system.
Using $G_{0,ij\sigma}^{-1}(\omega) = (\omega + \mu) \delta_{ij} - t_{ij}$, 
inserting the expansions into the C-DMFT self-consistency equation (\ref{eq:cdmft}), 
and expanding in powers of $\omega^{-1}$ once more, we immediately find:
\begin{equation}
 t'_{ij} = t_{ij} 
\end{equation}
for (correlated) sites $i$ and $j$ within the same cluster, and, in particular,
$\varepsilon' = \varepsilon$ where $\varepsilon' = t'_{ii}$ and $\varepsilon=t_{ii}$.
This means that, within the C-DMFT, consistency with respect to the particle number
is assured by setting the on-site energies within a cluster of the reference system 
to their ``physical'' values.

It goes without saying that a consistent determination of $\langle N \rangle$ 
requires an {\em exact} solution of the reference system which, in case of the C-DMFT,
is by no means trivial. 
An obvious idea is to replace the continuous bath by a few uncorrelated sites only
to allow for an application of the Lanczos method (see Ref.\ \onlinecite{ka.ci}, for
an example).
This, however, immediately implies that the self-consistency equation (\ref{eq:cdmft})
cannot be fulfilled exactly any longer and, strictly speaking, the determination of 
the particle number becomes inconsistent {\em unless} $\varepsilon'$ is treated as a
variational parameter within the SFT framework.


\end{document}